\documentclass[twocolumn,showpacs,preprintnumbers,amsmath,amssymb,
superscriptaddress,prb]{revtex4}

\usepackage{graphicx}
\usepackage{dcolumn}
\usepackage{bm}

\begin{document}

\title{Hopping conduction in strong electric fields: \\
Negative differential conductivity}

\author {A. V. Nenashev}
\email {nenashev@isp.nsc.ru}
\affiliation{Institute of Semiconductor Physics, 630090 Novosibirsk, Russia}
\affiliation{Novosibirsk State University, 630090 Novosibirsk, Russia}

\author {F. Jansson}
\affiliation{Graduate School of Materials Research,
\AA bo Akademi University,  20500 Turku, Finland}
\affiliation{Department of Physics and Center for Functional Materials,
\AA bo Akademi University, 20500 Turku, Finland}

\author {S. D. Baranovskii}
\affiliation{Department of Physics and Material Sciences Center, 
Philipps-University, 35032 Marburg, Germany}

\author {R. \"Osterbacka}
\affiliation{Department of Physics and Center for Functional Materials,
\AA bo Akademi University, 20500 Turku, Finland}

\author {A. V. Dvurechenskii}
\affiliation{Institute of Semiconductor Physics, 630090 Novosibirsk, Russia}
\affiliation{Novosibirsk State University, 630090 Novosibirsk, Russia}

\author{F. Gebhard}
\affiliation{Department of Physics and Material Sciences Center, 
Philipps-University, 35032 Marburg, Germany}

\date{\today}

\begin{abstract}
Effects of strong electric fields on hopping conductivity are
studied theoretically. Monte-Carlo computer simulations show that
the analytical theory of Nguyen and Shklovskii [Solid State
Commun. 38, 99 (1981)] provides an accurate description of hopping
transport in the limit of very high electric fields and low
concentrations of charge carriers as compared to the concentration
of localization sites and also at the relative concentration of
carriers equal to 0.5. At intermediate concentrations of carriers
between 0.1 and 0.5 computer simulations evidence essential
deviations from the results of the existing analytical theories.

The theory of Nguyen and Shklovskii also predicts a negative
differential hopping conductivity at high electric fields. Our
numerical calculations confirm this prediction qualitatively.
However the field dependence of the drift velocity of charge
carriers obtained numerically differs essentially from the one
predicted so far. Analytical theory is further developed so that
its agreement with numerical results is essentially improved.

\end{abstract}


\maketitle

\section{Introduction}
\label {sec-introduction}
Hopping conduction in solids governed by strong electric fields is
in the focus of intensive theoretical and  experimental study
since several decades [see, for instance, chapter 7 in
Ref.~\onlinecite{Bottger1985} and references therein].  In 
recent years, particular interest to this research area has been
caused by growing device applications of amorphous organic and
inorganic materials in which the incoherent hopping transitions of
charge carriers between spatially and energetically distributed
localized states dominate the optoelectronic phenomena [see, for
instance, Ref.~\onlinecite{Baranovski2006} and references
therein]. One of the mostly discussed topics is whether the
differential negative conductivity (NDC), i.e., the decreasing
conductivity with increasing electric field, is possible in the
hopping regime.  The discussion was, to much extent, provoked by
the reports on the apparent decrease of the drift mobility with
rising electric field at relatively high temperatures and low
field strengths in disordered organic
materials.\cite{Bassler2000,Pope1999,Bassler1993,
Borsenberger1993,Auweraer1994,Abkowitz1992,Peled1988,Schein1992}
This apparent decrease of the mobility with increasing electric
field was reported to be succeeded by the increase of the mobility
at higher field strengths. 
However, the self-consistent effective-medium theory for drift and 
diffusion at low electric fields\cite{Parris96} does not show any decrease 
of the mobility with increasing field. Furthermore, it has been shown
experimentally\cite{Hirao1995} and theoretically\cite{Cordes2001}
that the apparent decrease of the mobility with rising field at
low field strengths is an artifact. The experimental data were
obtained by the time-of-flight technique, in which charge carriers
are created close to one surface of a sample with a given
thickness $L$ and the transient time $\tau_\text{tr}$ is measured,
which is needed for charge carriers to reach the opposite surface
of the sample at a particular strength of the applied electric
field $F$.  Then the drift mobility is calculated as
$\mu=L/(\tau_\text{tr}F)$. However, at high temperatures and low
electric fields the current transients in the time-of-flight
experiments are determined mostly by diffusion of charge carriers
rather than by their drift. Therefore, using the drift formula one
strongly overestimates the mobility. It is the presence of the
field strength in the denominator that leads to the apparent
``increase'' of the mobility at decreasing
$F$.\cite{Hirao1995,Cordes2001} If one uses at low fields and high
temperatures the diffusion formulas instead of the drift ones,
then no decrease of the mobility with increasing field can be
claimed at low electric fields.\cite{Hirao1995,Cordes2001}

This result does not exclude, however, the possibility of the NDC in
the hopping regime. B\"ottger and Bryksin\cite{Bottger1979}  and
Shklovskii et al.\cite{NS,Levin1988} have suggested analytical
theories for the mobility and conductivity decreasing with
increasing electric field in various disordered materials.
Remarkably, this effect of the negative differential conductivity
is to be expected at high field strengths. This regime succeeds
the very strong increase of the mobility with rising
field,\cite{NS,Levin1988} and does not precede it at lower fields
as claimed on the basis of the drift
equations.\cite{Bassler2000,Pope1999,Bassler1993,Borsenberger1993,
Auweraer1994,Abkowitz1992,Peled1988,Schein1992}


The decreasing conductivity with increasing electric field at high field strengths
has been observed experimentally for hopping transport in lightly doped and
weakly compensated crystalline silicon.
\cite{Shklovskii99r,Shklovskii114r,Shklovskii116}
The hopping transport 
mode in such systems at low electric fields had been described theoretically in 
all detail, \cite{bible} which made these systems 
particularly attractive for studying the new non-Ohmic effects. 
Shklovskii et al.
\cite{NS,Shklovskii99r,Shklovskii114r,Shklovskii116} 
developed an analytical theory, which predicted the NDC effect in
the lightly doped weakly compensated semiconductors. The experimental
observations in lightly doped
and weakly compensated crystalline silicon appear in qualitative
agreement with his theoretical predictions.
Furthermore computer simulations of Levin et
al.\cite{Shklovskii76r} confirmed qualitatively the 
existence of the NDC effect, though no
quantitative comparison with the analytical theory \cite{NS} has been attempted.     
Recent interest in the NDC effect has been caused by its importance for 
construction of memory devices.
These devices typically contain conducting
particles embedded into a nonconductive material. For such
devices, made from both inorganic \cite{Simmons1967,Thurstans2002}
and organic
\cite{Bozano2004,Bozano2005,Majumdar2005,Verbakel2007,Baral2008}
materials, NDC and switching phenomena have been reported. Since
electrical conduction in the materials, which are currently being
tried for device applications, is dominated by hopping of charge
carriers, it is necessary to study the possibility of the NDC in
this regime in more detail.

In the present paper we report on the theoretical study of hopping
transport in high electric fields.  In Section \ref{sec-model} we
describe the theoretical model and briefly outline the analytical
approach suggested by Nguyen and Shklovskii\cite{NS} for the limit
of extremely high electric fields. In Section \ref{sec-mc-inf} we
present our results obtained by straightforward Monte Carlo
computer simulations and show the range of applicability for the
analytical theory of Nguyen and Shklovskii. In Section
\ref{sec-our-theory} we further develop the analytical theory whereby we
improve its agreement with the results of computer simulations.
In particular, the analytical dependence of the drift mobility on
the concentration of charge carriers comes in better agreement
with the simulation results. Section \ref{sec-finite} is dedicated
to the NDC. A new numerical algorithm has been developed to study
the NDC effect theoretically. Numerical results obtained in the
framework of this algorithm confirm qualitatively the conclusion
of Nguyen and Shklovskii on the possibility of the NDC in the
hopping regime. However, the field dependence of the drift velocity
of charge carriers obtained numerically differs essentially from
the one predicted so far. \cite{NS} We suggest in Section
\ref{sec-finite} a further development of the analytical theory,
improving essentially its agreement with numerical results.
Concluding remarks are gathered in Section \ref{sec-conclusions}.

\section{Model and theoretical background}
\label {sec-model}

Aiming to clarify whether the NDC effect is inherent for the
hopping transport regime, we consider first, following Nguyen and
Shklovskii,\cite{NS} the simplest possible model---a
three-dimensional array of isoenergetic sites with a random
spatial distribution with the concentration $N$. Each site can be either 
empty or occupied by
a single electron. Energies of electrons are equal on all sites so
that no energy disorder and no electron-electron interactions
between different sites are taken into account. 
Only in the final part of Section \ref{sec-finite} we study the 
effect of the energy disorder on the NDC.
An electric field
$\mathbf F=(-F,0,0)$ is put along the negative direction of the
axis $X$, so that the drift velocity of the negatively charged
electrons is directed along the $X$ axis. Conduction takes place
due to tunnelling hops of electrons between the localization
sites. The rate $\Gamma_{ij}$ for an electron hop from site $i$
to site $j$ is determined as
\begin{equation}
\label{eq-rate}
 \Gamma_{ij}=\Gamma_0 \exp\left(-\frac{2d_{ij}}{a}\right) 
f\left(\frac{eF(x_j-x_i)}{kT}\right) n_i (1-n_j),
\end{equation}
where $d_{ij}$ is the distance between the sites, $a$ is the
localization length, $e$ is the elementary charge, $k$ is  the
Boltzmann constant, $T$ is the temperature, and $n_i,n_j$ are the
occupation factors of the sites, $(n_i,n_j\in\{0;1\})$. The
function $f$ is related to energy the gain or the energy loss
during the jump:
\[
 f(\alpha)= \begin{cases}
             1,            & \text{if } \alpha>0, \\
             \exp(\alpha), & \text{if } \alpha<0.
            \end{cases}
\]
In the limit of infinite electric field, the factor 
$f\left[eF(x_j-x_i)/kT\right]$ reduces to the Heaviside's 
function $\theta(x_j-x_i)$.

Below we will assume that $\Gamma_0=e=1$.
As a measure of length, we introduce the typical distance 
between the neighboring sites $R=N^{-1/3}$.

\begin{figure}
\includegraphics[width=3.4in]{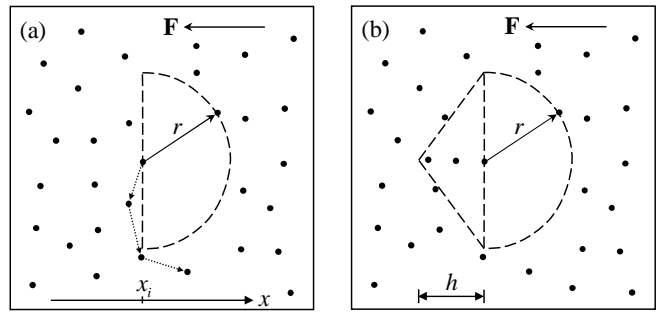}
\caption{The shape of optimal traps for the infinite~(a) and a
finite~(b) electric fields. The dotted path in (a) is forbidden at infinite 
fields but provides an escape route at finite fields} \label {fig-traps}
\end{figure}

Using this simple model, Nguyen and Shklovskii\cite{NS} have shown
analytically that the effect of the NDC is inherent for hopping
transport. Let us consider briefly their arguments starting from
the case of infinitely high fields $F$ and extremely small
electron concentrations $n_e$. Under such circumstances each
electron can be treated independently from the others and
electrons can move only toward the increasing values of their $x$
coordinate. In Fig.~\ref{fig-traps} this is the direction to the
right. At each jump, an electron moves along the axis $X$ to a
distance $\simeq R$, so that its drift velocity can be estimated as
$v\simeq R/\overline{\tau}$, where $\overline{\tau}$ is an average
time between jumps (dwell time). A dwell time $\tau_i$ for hopping
from the  site $i$ is of the order of $\exp(2r_i/a)$, where $r_i$
is the distance from site $i$ to its nearest neighbor ``to the
right'', i.e.\ with co-ordinate $x$ larger than $x_i$. In other
words, $r_i$ is the maximum radius of a hemisphere centered at the
 site $i$ that does not contain any other sites
(Fig.~\ref{fig-traps}(a)). If $r_i$ is much larger than $R$ (the
typical distance between the neighboring sites),
such an empty hemisphere can be considered as a trap for
electrons. The contribution of traps with radii in the range
$[r,r+dr]$ to the average dwell time $\overline{\tau}$ is
proportional to $\tau(r)=\exp(2r/a)$, and also to the probability
of the corresponding configuration of sites, $p(r)dr=2\pi Nr^2
\exp(-2\pi Nr^3/3)dr$:
\begin{equation}
\label{eq-shklovskii4}
 \overline{\tau} = \int\limits_0^\infty \tau(r)p(r)dr 
= \int\limits_0^\infty 2\pi Nr^2 \exp\left(\frac{2r}{a}-
\frac{2\pi N}{3}r^3\right)dr.
\end{equation}
This integral is easy to evaluate, taking into account that the
integrand has a sharp maximum at $r=r_m\equiv 1/\sqrt{\pi Na}$.
Consequently one obtains for the current density $j = n_e v$
\begin{equation}
\label{eq-shklovskii5}
 j_{F\rightarrow\infty,n_e\rightarrow0} \simeq \frac{n_eR}{\overline\tau} 
\simeq n_e (a^3R)^{1/4} \exp\left(-\frac{4}{3\sqrt\pi}
 \left(\frac{R}{a}\right)^\frac32\right).
\end{equation}

Therefore, one can conclude that in the limit
$F\rightarrow\infty,n_e\rightarrow0$ the current is determined by
hemispherical traps (Fig.~\ref{fig-traps}a)  with an ``optimal''
radius $r_m=1/\sqrt{\pi Na}$.

In the case of finite electric fields, a hemispherical trap is not
an efficient one, because an electron has a possibility to move in
the energetically unfavored directions, and thus to escape the
trap (for example, along the dotted arrows in
Fig.~\ref{fig-traps}(a)). According to Nguyen and
Shklovskii,\cite{NS} an ``optimal'' trap for an electron in large
though finite electric fields $F$ consists of a hemisphere to the
right and of a cone to the left of the site on which an electron
is captured, with a chain of sites along the $X$ axis that
provides an easy path into the trap (Fig.~\ref{fig-traps}b). The
height $h$ of the cone is chosen so that it is equally hard to
escape the trap in all directions taking the chain along the $X$
axes into account: $h=2rkT/Fa$. Therefore, the smaller is the
field, the larger is the volume of a trap with the same dwell
time, and consequently the smaller is the probability $p(r)$ of
finding such a trap. It means that the average dwell time
$\overline{\tau} = \int_0^\infty \tau(r)p(r)dr$ decreases with
decreasing field strength, and concomitantly the current density
$j \simeq n_eR/\overline\tau$ increases with decreasing field. This
is the essence of the physical mechanism that causes the NDC
effect.\cite{NS} To obtain an expression for the current density,
one can substitute the volume of the trap shown in
Fig.~\ref{fig-traps}b, $V_\text{trap} = (1+kT/Fa) 2\pi r^3/3$,
instead of the hemispherical trap volume, $2\pi r^3/3$, into the
integral (\ref{eq-shklovskii4}). The result reads
\begin{equation}
\label{eq-shklovskii10}
 j_{n_e\rightarrow0} \simeq  n_e (a^3R)^{1/4} 
\exp\left[-\frac{4}{3\sqrt\pi} \left(\frac{R}{a}\right)^\frac32
\left( 1+\frac{kT}{Fa} \right)^{-\frac12}\right].
\end{equation}
This is the mathematical expression for the NDC. The approach
leading to this expression is applicable only for fields $F\gg
kT/R$. In smaller fields, the assumption that almost every jump is
directed along the axis $X$ is violated. Therefore one should
expect that Eq.~(\ref{eq-shklovskii10}) overestimates the current
density for $F\simeq kT/R$.

Equations (\ref{eq-shklovskii5}) and (\ref{eq-shklovskii10}) are
valid only if the concentration of electrons $n_e$ is small as
compared to the concentration of ``optimal'' traps, $n_m =
N\exp[-NV_\text{trap}(r_m)]$. In the opposite case, $n_m \ll n_e \ll
N$, the ``optimal'' traps are almost always occupied and play a
negligible role. In such a case the most important traps, which
determine the drift velocity of electrons, are the ones whose
concentration is equal to $n_e$. One can estimate the electron drift
velocity as $v = 1/ \tau(r_n) n_e S$, where $r_n$ is a radius of
the most important traps, $\tau(r_n)$ is their dwell time, $S$ is
their capture crosssection. Assuming $S\simeq R^2\equiv N^{-2/3}$, 
one obtains for
the current flow $j = n_e v \simeq N^{2/3}\tau(r_n)^{-1}$. For an
infinitely large field, the radius $r_n$ is defined via
\[
 n_e = N\exp(-NV_\text{trap}(r_n)) \equiv 
N\exp\left(-\frac{2\pi N}{3}r_n^3\right),
\]
that gives $r_n=R\left( \frac{3}{2\pi} \log\frac{N}{n_e}
\right)^{1/3}$. Consequently, the current density is\cite{NS}
\begin{equation}
\label{eq-shklovskii7}
  j_{ \begin{array}{l} \scriptstyle F\rightarrow\infty, \\ 
\scriptstyle n_m \ll n_e \ll 1 \end{array} } 
\simeq \frac{N^{2/3}}{\tau(r_n)} \simeq N^\frac23\exp\left[ -\frac{2}{a} 
\left( \frac{3}{2\pi} \log\frac{N}{n_e} \right)^{\frac13} \right].
\end{equation}

The corresponding expression for the concentration range  $n_m \ll
n_e \ll N$ in the case of finite electric fields was also obtained
by Nguyen and Shklovskii [see Eq.~(11) in Ref.~\onlinecite{NS}].

The case of almost filled sites, $n_e \approx N$, is similar to
the case of almost empty sites, $n_e \approx 0$ due to
electron-hole  symmetry. The current density is a symmetrical
function of the electron concentration: $j(n_e)=j(N-n_e)$.

Nguyen and Shklovskii\cite{NS} also emphasized that a special
consideration is needed for the case of half-filled system,
$n_e=N/2$. They have shown that the concept of directed
percolation can be used to obtain the current density at
infinitely high electric fields. In the half-filled system the
trapping of electrons does not play any role, because (due to the
electron-hole symmetry) it does not change the electron concentration
on the infinite cluster which is responsible for the current.
Current is determined by electron jumps to distances
$d\in[r_c^d,r_c^d+a/2]$, where $r_c^d$ is the percolation
threshold of a directed percolation problem. The number of pairs
of sites with distances $d\in[r_c^d,r_c^d+a/2]$ in the infinite
cluster per unit area is $1/L_\perp^2$, where $L_\perp = R\,
(2r_c^d/a)^{\nu_\perp}$ is a transversal correlation length of the
percolation cluster, and $\nu_\perp$ is a critical index.\cite{NS}
The current density is equal to\cite{NS}
\begin{equation}
\label{eq-shklovskii3}
  j_{ \begin{array}{l} \scriptstyle F\rightarrow\infty, \\ \scriptstyle n_e=1/2 
\end{array} } \simeq
  \frac{1}{L_\perp^2 \tau(r_c^d)} = N^\frac23 
\left( \frac{a}{2r_c^d} \right)^{2\nu_\perp} 
\exp\left( -\frac{2r_c^d}{a} \right).
\end{equation}
Nguyen and Shklovskii\cite{NS} have also obtained the value of the
percolation threshold $r_c^d = (0.93\pm0.01)\,R$ and that of the
correlation length index $\nu = 1.2\pm0.1$.

The above arguments of Nguyen and Shklovskii\cite{NS} provide an
analytical theory of non-Ohmic hopping conduction, based on the
concept of the trapping-determined transport. The theory is valid
for the case of large electric fields in two concentration ranges:
$n_e \ll n_m$ and $n_m \ll n_e \ll N$. Most remarkably, this theory
predicts the effect of the NDC. Also a theory for the case of the
half-filled system $(n_e=N/2)$ for infinitely high electric fields
$(F\rightarrow\infty)$ has been suggested based on the
directed-percolation-approach.\cite {NS}

Below we present our numerical study of the field-dependent
hopping conductivity. It shows the range of validity for the
analytical theory of Nguyen and Shklovskii.\cite{NS} Furthermore,
the analytical theory is developed below in order to improve the
agreement between the analytical and numerical results.

\section {Monte Carlo simulations for infinitely high fields}
\label {sec-mc-inf} In order to calculate the electron drift
velocity and the current density at high fields, we used a Monte
Carlo approach. In the limit of infinitely high fields the
direction of the electron motion is prescribed. Therefore it was
possible to simulate by a Monte Carlo algorithm the motion of an
electron in an infinite medium along the field direction and
therefore to avoid any size effects. Without loosing generality
one can restrict the maximal length of electron transitions
involved into the algorithm by a reasonably large value
$d_\mathrm{max}$. In order to simulate the $k$-th Monte Carlo step
in the electron motion, one has to store information only about
sites inside a layer $x_k<x<x_k+d_\mathrm{max}$, where $x_k$ is the
electron coordinate before the $k$-th step. We have chosen
$d_\mathrm{max}=3R$, which provides a possibility to hop to $2\pi
d_\mathrm{max}^{3}N/3\simeq57$ neighbors in average. For all sets of
parameters used in the simulation the size of the optimal trap $r_{m}$
considered by Nguyen and Shklovskii was essentially less than
$d_\mathrm{max}$. Therefore, the restriction imposed by
$d_\mathrm{max}$ did not lead to any loss of generality. Before
making the next step, the computer can forget all the information
about sites in the layer $x_k<x<x_{k+1}$, but it has to get
information about new sites in a layer
$x_k+d_\mathrm{max}<x<x_{k+1}+d_\mathrm{max}$. As these ``new''
sites did not affect the calculation at all previous steps, they
can be created at random. Therefore, each Monte Carlo step
includes not only the choice of a jump, but also a generation of
some ``new'' sites and deleting some ``old'' sites. To make their
number finite, one should restrict the system size in the
directions perpendicular to the field, i.e.\ in the plane $YZ$.
A calculation domain
$0<y<120R$, $0<z<120R$ with periodical boundary conditions in the
plane $YZ$ was used. The motion of a single electron was simulated
within the described algorithm in order to evaluate the drift
velocity in the limit $n_e\rightarrow0$.

For finite electron concentrations, we perform simulations in a
cubic domain with size $60R \times 60R \times60R$ and with periodic
boundary conditions for all three axes. The rates of all possible
jumps are calculated before starting the Monte Carlo steps, but
without the factor $n_i (1-n_j)$ related to occupation. This
factor determines which jumps are allowed and which are forbidden.
Information about allowed and forbidden jumps is updated at each
step. We used a binary-tree data structure for storing the jump
rates that gives the possibility to ``switch on'' and ``off''
jumps efficiently.

\begin{figure}
\includegraphics[width=3.4in]{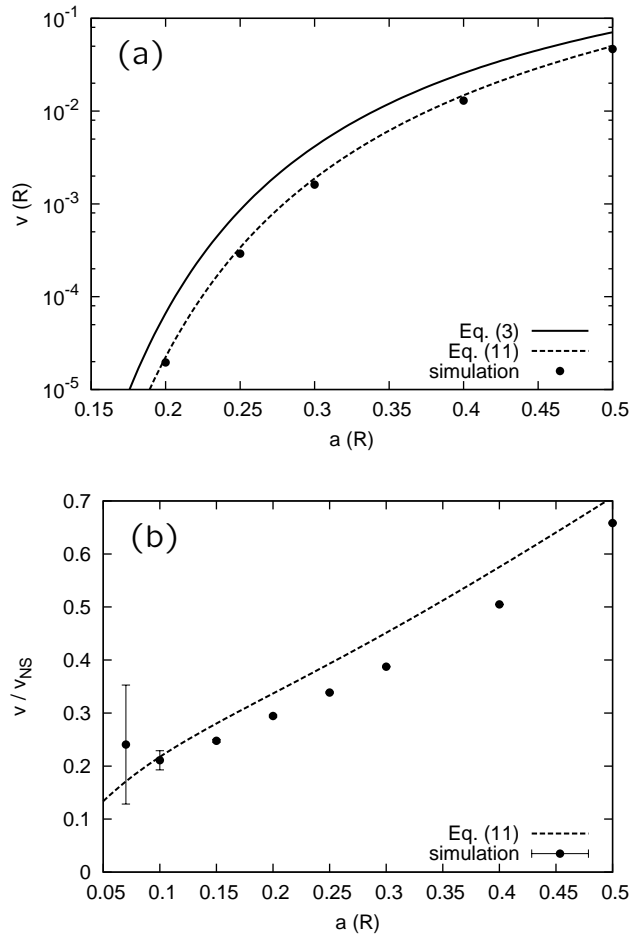}
\caption{(a) Drift velocity $v$ in the limit of infinitely large
electric field and small electron concentration, as a function of
localization length $a$; (b) the same data divided by the velocity
value predicted in Ref.~\onlinecite{NS}
(Eq.~(\ref{eq-shklovskii5})).} \label {fig-zero-n}
\end{figure}

A routine Monte Carlo procedure has been used. In each Monte Carlo
step the final site for electron hops was calculated via the
probabilities proportional to the hopping rates to different sites
and the time $\Delta t$ spent to jump was calculated via the
reciprocal of the sum of rates of all possible jumps. Hops from an
occupied site were possible to any empty one in the direction of
increasing coordinate $x$ with the restriction that the hop
distance is less than $d_\mathrm{max}\equiv3R$. At each hopping
event the increment $\Delta x$ in electron $x$-coordinate is
calculated. An outcome of the simulation is either an average
velocity of an electron,
\[
 v=\sum \Delta x / \sum \Delta t,
\]
in the case of single electron hopping, or a flow of electrons,
\[
 j=\frac1\Omega \sum \Delta x / \sum \Delta t,
\]
in the case of finite electron concentrations (where $\Omega$ is
the volume of the calculation domain). The summation was carried
out over all sequential Monte Carlo steps. For simulations of the
behavior of a single electron in an empty system, $10^{7}$ Monte
Carlo steps were used for $a\geqslant0.2R$,
$10^{8}$ steps for $a=0.10R$ and $0.15R$ and $10^{9}$ steps
for $a=0.07R$. As a result, for $a\geqslant0.15R$ convergence was
not worse than $1\%$.

For finite electron concentration, $5\cdot10^{7}$ Monte Carlo
steps were used, this gave a convergence not worse than $1\%$
for a given realization. At $a\leqslant0.1R$, there were sometimes
essential differences between current densities in different
realizations. The scatter is shown by error bars in the figures.

Simulation results for the electron drift velocity $v=j/n_e$  in
the limit $n_e\rightarrow0$ are shown in Fig.~\ref{fig-zero-n}(a) by
dots as a function of the localization length. The analytical
result of Nguyen and Shklovskii (Eq.~(\ref{eq-shklovskii5})) is
shown by the solid line. One can see that
Eq.~(\ref{eq-shklovskii5}) correctly describes the dependence of
the drift velocity on the localization length and, furthermore, it
correctly estimates the magnitude of the velocity. The concept of
Nguyen and Shklovskii on the hopping drift velocity controlled by
hemispherical traps is herewith confirmed. However, there is some
deviation of the simulation data from the analytical results. To
make this deviation more transparent, we plot the ratio of the
simulated drift velocity to its analytical prediction
(Eq.~(\ref{eq-shklovskii5})) in Fig.~\ref{fig-zero-n}(b). It is seen
that Eq.~(\ref{eq-shklovskii5}) overestimates the electron
velocity by a factor of two to five. In Section~\ref{sec-our-theory}A, some
reasons for this mismatch will be considered. The analytical
theory is further developed there to give a better agreement with
the simulation data. The result of the improved theory for the
drift velocity (Eq.~(\ref{eq-v-improved})) is also shown in
Fig.~\ref{fig-zero-n} by the dashed line.

\begin{figure}
\includegraphics[width=3.4in]{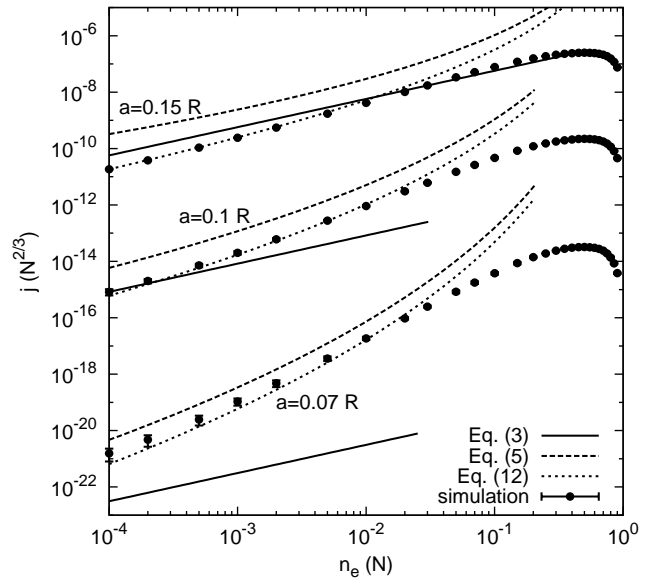}
\caption{Current density $j$ as a function of electron
concentration $n_e$ for values of the localization length $0.2R$,
$0.1R$, and $0.07R$ (from top to bottom). The electric field is
infinitely large.} \label {fig-n-dep}
\end{figure}

\begin{figure}
\includegraphics[width=3.4in]{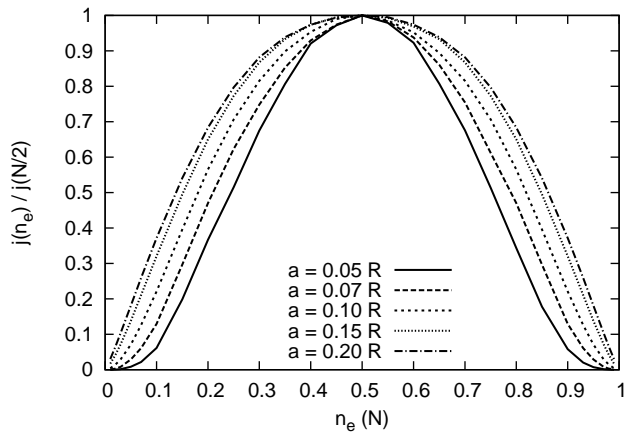}
\caption{Current density $j$ as a function of electron
concentration $n_e$, normalized on its maximum value $j(N/2)$, at
different  localization lengths. The electric field is infinitely
large.} \label {fig-near-half}
\end{figure}

The dependence of the current density on the electron
concentration is shown in Figs.~\ref{fig-n-dep} and
\ref{fig-near-half}.  Fig.~\ref{fig-n-dep} shows this dependence
in a wide concentration range, in comparison with the analytical
results for small ($n_e\ll n_m$, Eq.~(\ref{eq-shklovskii5}), solid
line) and intermediate ($n_m\ll n_e\ll N$,
Eq.~(\ref{eq-shklovskii7}), dashed line) concentrations. One can
see that the simulated concentration dependence can be roughly
divided into three parts: for very low concentrations ($n_e<n_m$)
the dependence is linear,\footnote{This linear part in seen in
Fig.~\ref{fig-n-dep} only for $a=0.15R$, because for $a=0.1R$ and
$0.07R$ the value $n_m$ is less than $10^{-4}N$.} in accordance with
Eq.~(\ref{eq-shklovskii5}); then, for $n_m<n_e<0.03N$, it becomes
superlinear, as described by Eq.~(\ref{eq-shklovskii7}); and
finally, for $n_e>0.03N$, this dependence is sublinear and is not
described by the theory based on the transport controlled by
traps. In Section~\ref{sec-our-theory}B, we will present an
analytical approach valid for the range of parameters covering the
ranges of applicability of Eqs.~(\ref{eq-shklovskii5}) and
(\ref{eq-shklovskii7}). The result of this developed approach is
Eq.~(\ref{eq-n-integral}) shown by dashed lines in
Fig.~\ref{fig-n-dep}.\footnote{The integral in
Eq.~(\ref{eq-n-integral}) is evaluated numerically.} One can see
that it provides an accurate description of the current density
for any concentration less than $0.03N$.

For $n_e>0.03N$, the simulated values of the current density are
smaller than those predicted by the analytical theory due to the
following reason. At sufficiently large electron concentrations,
the conducting paths are not almost empty, as is assumed in the
theory. Moreover, there are ``bottlenecks'' for the current, where
the electron concentration is much larger than the mean
concentration $n_e$. In these places, the factor of $(1-n_j)$ in
Eq.~(\ref{eq-rate}) turns out to be important, and due to this
factor the current density is suppressed.

In Fig.~\ref{fig-near-half}, the simulation results are shown for
the whole range of carrier concentrations. For convenience, all
values of current density are divided by the maximum value for the
given localization length. For large localization lengths
($a\geq0.15R$), the concentration dependence of the current
density~$j$ obeys approximately a parabolic law: $j(n_e) \sim
n_e(N-n_e)$ at concentrations in the vicinity of the half filling.
One can interpret this behavior in terms of the hopping
rates, namely, by substituting the mean occupancy $n_e/N$ instead
of $n_i$ and $n_j$ into Eq.~(\ref{eq-rate}). Concomitantly, one
obtains that the contribution of each pair of sites is
proportional to $n_e(N-n_e)$. The same concentration dependence is
expected then for the current density.

\begin{figure}
\includegraphics[width=3.4in]{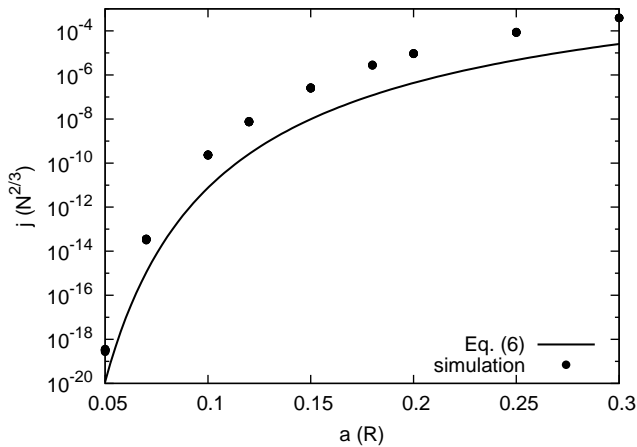}
\caption{Current density $j$ as a function of localization length
$a$ for the half-filled system $(n_e=N/2)$. The electric field is
infinitely large.} \label {fig-half-fill}
\end{figure}

Fig.~\ref{fig-half-fill} shows the simulated dependence of the
current density on the localization length (dots) in comparison
with  the analytical theory based on the concept of directed
percolation (Eq.~(\ref{eq-shklovskii3}), solid line) for
$n_e=N/2$. Apparently, the theory of Nguyen and
Shklovskii\cite{NS} correctly describes this dependence within the
range of current densities of almost 15 orders of magnitude.
However, the theory underestimates the magnitude of the current
density by approximately a factor of 30. Further research is necessary
to clarify the reasons of this discrepancy.

\section {Analytical theory for infinitely high fields}
\label{sec-our-theory}

Our numerical studies show that although the analytical
description of hopping conduction in very strong electric field by
Nguyen and Shklovskii  is qualitatively correct the quantitative
predictions differ sometimes by more than an order of magnitude
from the numerical results. In this Section, we show how to
improve the accuracy of the analytical theory.

\subsection{Limit of $n_e \rightarrow 0$}

For low electron concentrations $n_e \ll n_m$, where
$n_m=N\exp[-\frac{2\pi}{3}(R/\pi a)^{3/2}]$ is the concentration of
the ``optimal'' traps,  the prediction of Ref.~\onlinecite{NS} for
the electron drift velocity is expressed by
Eq.~(\ref{eq-shklovskii5}). Now we discuss several corrections to
this equation.

1) There is a numeric factor of $(4\pi)^{1/4} \approx 1.88$ in
$\overline\tau$, arising from the evaluation of the
integral~(\ref{eq-shklovskii4}) that should be taken into account.
It gives a factor of $(4\pi)^{-1/4}$ for the drift velocity.

2) The mean electron displacement along the $X$ axis,
$\langle\Delta x\rangle$, is taken equal to $R$ in Ref.~\onlinecite{NS}.
We performed Monte Carlo calculations for
$\langle\Delta x\rangle$ as a function of the localization length
$a$ and get the following fitting expression:
\begin{equation}
\label{eq-deltax}
  \langle\Delta x\rangle = R\,(0.385+0.45a^2/R^2)
\end{equation}
(the accuracy of fitting is not worse than 0.3~\% in the range
$0.05\leq a/R\leq0.2$). Therefore, the drift  velocity
$v=\langle\Delta x\rangle/\overline\tau$ gets an additional factor
equal approximately to $0.5$.

3) The dwell time $\tau(r)$ of a trap with a radius $r$ is in fact
somewhat less than the value $\exp(2r/a)$ used in
Ref.~\onlinecite{NS} because an electron can escape the trap by
moving not only to the nearest site to the right, but also to a
more distant site. A contribution $\Gamma_1$ of these distant
sites to the escaping rate is
\[
 \Gamma_1 = \int\limits_r^\infty e^{-2r_1/a} 2\pi Nr_1^2 dr_1 
= e^{-2r/a} \pi Na \left(r^2\!+\!ar\!+\!\frac{a^2}{2}\right).
\]
Then, the dwell time $\tau(r)$ is a reciprocal of the sum
$\Gamma_0+\Gamma_1$, where $\Gamma_0=\exp(-2r/a)$ is the rate of a
jump to the nearest neighbor:
\begin{equation}
\label{eq-tau}
  \tau(r) = \frac{1}{\Gamma_0+\Gamma_1} 
= \frac{\exp(2r/a)}{1+\pi Na (r^2+ar+a^2/2)}.
\end{equation}
For $r=r_m\equiv (\pi Na)^{-1/2}$, $\tau(r)$ is approximately half of
 $\exp(2r/a)$, that results in a factor of two in
the  drift velocity.

\begin{figure}
\includegraphics[width=3.4in]{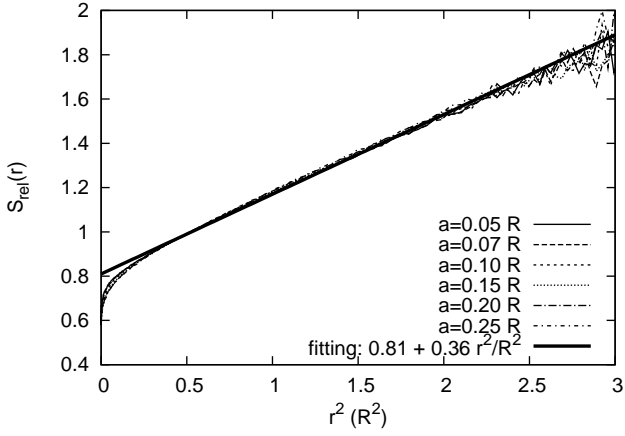}
\caption{Relative capture crosssection as a function of squared trap radius.}
\label{fig-cross}
\end{figure}

4) The geometrical crosssections of larger traps have larger
capture crosssections for electrons than the smaller ones. This
results in different probabilities for carriers to be captured
by traps with different radii. The probability $\tilde p(r)dr$
that the next visited site will be a trap with radius in the range
$(r,r+dr)$ is
\[
 \tilde p(r)dr = \frac{S(r)}{\langle S\rangle}p(r)dr,
\]
where $p(r)=2\pi Nr^2\exp(-2\pi Nr^3/3)$, $S(r)$ is a capture 
crosssection of a trap with radius $r$, and $\langle S\rangle 
= \int S(r)p(r)dr$ is a mean crosssection. Below we will use a 
notation $S_\text{rel}(r)$ for a ``relative crosssection'' 
$S(r)/\langle S\rangle$. Then, instead of Eq.~(\ref{eq-shklovskii4}) 
we get
\begin{equation}
\label{eq-dwelltime-s}
  \overline{\tau} = \int\limits_0^\infty \tau(r)S_\text{rel}(r)p(r)dr.
\end{equation}
We calculated the relative crosssections with Monte Carlo method
as ratios $N_{tr}[r,r\!+\!\Delta r]/(N_j p(r)\Delta r)$, where
$N_{tr}[r,r\!+\!\Delta r]$ is a number of traps with radii in the
specified range visited by an electron, and $N_j$ is a total
number of electron jumps. We used $N_j=10^8$ and $\Delta r=0.01R$.
The results are presented in Fig.~\ref{fig-cross}. The relative
crosssection is almost independent of the localization length for
$r>0.3R$. Its dependence on the trap radius is described by the
quadratic function:
\begin{equation}
\label{eq-s-rel}
  S_\text{rel}(r)=0.81+0.36\,r^2/R^2.
\end{equation}
For the ``optimal'' traps with $r=r_m\equiv(\pi Na)^{-1/2}$ we get
$S_\text{rel}(r_m)\sim a^{-1}$ in the limit $a\rightarrow0$. According
to Eq.~(\ref{eq-dwelltime-s}), it results in a factor of $\sim
a^{-1}$ for the mean dwell time $\overline\tau$ and, consequently,
in a factor of $\sim a$ for the drift velocity.

Now we can improve Eq.~(\ref{eq-shklovskii5}) of Nguyen and
Shklovskii, starting from Eq.~(\ref{eq-dwelltime-s}). Since the
integrand has a sharp maximum at $r_m=(\pi Na)^{-1/2}$, we can
estimate the integral approximately as
\[
 \overline{\tau} \approx \tau(r_m)\, S_\text{rel}(r_m)\, p(r_m)\, 
(\pi R^3a/4)^{1/4}N^{-1}.
\]
Then, using Eqs.~(\ref{eq-deltax}), (\ref{eq-tau}), and (\ref{eq-s-rel}), 
we get the following expression for the drift velocity
$v={\langle\Delta x\rangle}/{\overline\tau}$:
\[
 v  \approx
     \frac{(0.85\!+\!0.45\frac{a^2}{R^2})\left[1\!+\!\pi Na 
\left(\frac1{\pi Na}\!+\!\sqrt{\frac a{\pi N}}\!+\!\frac{a^2}2\right)\right]}
     {(4\pi)^{1/4} \left(0.81+0.36\frac{R}{\pi a}\right)} \,v_{\mathrm NS}.
\]
Here $v_{\mathrm NS}=j/n_e$ is the drift velocity corresponding to
Eq.~(\ref{eq-shklovskii5}). Finally, the latter expression can be
fitted (with accuracy of about 3~\% for $a\leq 0.2R$) by a simple formula,
\begin{equation}
\label{eq-v-improved}
  v \approx \frac{a}{R} \left(1.4+2.1\, e^{-10\,a/R}\right) v_{\mathrm NS}.
\end{equation}
This expression is to be considered as a corrected analytical form
for the drift velocity at infinitely high fields in the limit of
small electron concentration.

A comparison of Eq.~(\ref{eq-v-improved}) with the values of the
drift velocity obtained by the Monte Carlo method is shown in
Fig.~\ref{fig-zero-n}.  The difference between the analytical and
simulated results does not exceed 20~\%. We believe that the main
source of this small difference is some inaccuracy in determining
$\tau(r)$ by Eq.~(\ref{eq-tau}). In fact, for a given trap radius
there is some distribution of the dwell times. The quantity
$\tau(r)$ contributing to Eq.~(\ref{eq-dwelltime-s}) is the mean
dwell time for radius $r$. However, Eq.~(\ref{eq-tau}) gives the
reciprocal value of the mean escaping rate that is slightly
smaller than $\tau(r)$. For this reason, Eq.~(\ref{eq-v-improved})
can slightly overestimate the drift velocity.

\subsection{Finite electron concentration}

Let us now try to improve the analytical approach at finite,
though small electron concentration, $n_e\ll N/2$. In this case,
electron flow can be considered as a homogeneous one on the scale
of distances between the traps that determine the
transport. Hence one can express the frequency $\nu_{in}$ of
electron capture by a trap as $\nu_{in}=jS(1-\bar{n})$, where $j$
is the current density, $S$ is the trap capture crosssection, and
$\bar{n}$ is its mean occupancy. Under the steady-state
conditions, $\nu_{in}=\nu_{out}$, where $\nu_{out}=\bar{n}/\tau$
is a frequency of emission of electrons from the trap, $\tau$ is a
dwell time. From this equation one can get $\bar{n}$:
\[
 \bar{n} = \frac{1}{1+(jS\tau)^{-1}}.
\]
Since in a snapshot of the system almost all electrons are
captured by traps, the total electron concentration $n_e$ is
\begin{equation}
\label{eq-n-integral}
  n_e = \int\limits_0^\infty \bar{n}(r)p(r)dr 
= \int\limits_0^\infty \frac{p(r)dr}{1+(jS(r)\tau(r))^{-1}}.
\end{equation}
The dwell time $\tau(r)$ can be estimated by Eq.~(\ref{eq-tau}).
In order to find the crosssection $S(r)$, one should note that
the mean crosssection $\langle S\rangle$ is equal to $1/ N \langle
\Delta x\rangle$. Consequently,
\begin{equation}
\label{eq-S}
  S(r) = \frac{S_\text{rel}(r)}{N\langle \Delta x\rangle} 
= R^2\,\frac{0.81\,R^2+0.36\,r^2}{0.385\,R^2+0.45\,a^2}.
\end{equation}

Equation~(\ref{eq-n-integral}) with $\tau(r)$ and $S(r)$
determined by Eqs.~(\ref{eq-tau}) and (\ref{eq-S}), respectively,
gives a functional dependence between the electron concentration
and the current density for any $n_e\ll N/2$. Fig.~\ref{fig-n-dep}
evidences a good agreement between Eq.~(\ref{eq-n-integral}) and
the Monte Carlo calculations for $n_e\leq0.03\,N$.

Although there is probably no simple way to resolve
Eq.~(\ref{eq-n-integral}) with respect to $j$ analytically in the
general case, it is  possible to simplify this equation in some
limiting cases. For small $n_e$ and $j$ $(n_e\ll n_m)$, the unity
term in the denominator of Eq.~(\ref{eq-n-integral}) can be
dropped, and we get $n_e=j/v$, where the drift velocity $v=\langle
\Delta x\rangle / \bar{\tau}$ is determined by
Eq.~(\ref{eq-v-improved}). In the opposite limit $(n_e\gg n_m)$,
one can evaluate Eq.~(\ref{eq-n-integral}) as
\[
 n_e \approx \int\limits_{r_n}^\infty p(r)dr 
= N\exp\left( -\frac{2\pi Nr_n^3}{3} \right),
\]
where a cutting parameter $r_n$ is given by condition $jS(r_n)\tau(r_n)=1$. 
Therefore,
\begin{equation}
\label{eq-j-upper-limit}
  j = \frac{1}{S(r_n)\tau(r_n)}
\end{equation}
with
\begin{equation}
\label{eq-r-n}
  r_n = R \left( \frac{3}{2\pi} \log\frac{N}{n_e} \right)^{1/3}.
\end{equation}
Equation~(\ref{eq-j-upper-limit}), with parameters determined by
Eqs.~(\ref{eq-tau}), (\ref{eq-S}) and (\ref{eq-r-n}) is the
improved version of Eq.~(\ref{eq-shklovskii7}) by Nguyen and
Shklovskii for the concentration range $n_m \ll n_e \ll N/2$.

\section {Hopping transport at finite electric fields}
\label{sec-finite} So far we have considered the limiting case of
infinitely high electric fields. Let us now turn to the field
dependence of the charge carriers velocity in order to reveal the
possibility of the negative differential conductivity
predicted by Nguyen and Shklovskii.\cite{NS}
Eq.~(\ref{eq-shklovskii10})
predicts a decreasing drift velocity with increasing electric
field, provided the field is strong enough. On the other
hand, for very small fields, Ohmic transport can be expected,
i.e., the drift velocity should depend linearly on the field. In
order to simulate hopping transport at finite electric fields, we
 solved a system of balance equations instead of using a direct MC
simulation. In the following subsection A we describe the details
of the numerical procedure, while the results are presented in 
subsection B.

\subsection {Balance equation method}
We consider a cubic system (side length $L$) with randomly placed sites.
Periodic boundary conditions are used in all directions.
The balance equation for the occupation probability $p_{i}$ of a
site $i$ has the form 
\cite{Yu2000,Yu2001,Pasveer2005,Cottaar2006,Jansson2008,Jansson2008b}
\begin{equation}
\label{eq-balance}
 \sum_{j \ne i} p_i \Gamma_{ij} (1-p_j) =
 \sum_{j \ne i} p_j \Gamma_{ji} (1-p_i).
\end{equation}

If all occupation probabilities $p_i$ are small,
i.e.\ the charge carrier concentration is low,
the balance equation can be linearized:
\begin{equation}
\label{eq-balance-lin}
 \sum_{j \ne i} p_i \Gamma_{ij} =
 \sum_{j \ne i} p_j \Gamma_{ji}.
\end{equation}

These equations are solved by defining
\begin{equation}
\mathbf{p} =
\left( \begin{array}{c}
p_1\\
p_2\\
p_3\\
\vdots \\
\end{array} \right)
\ \mathrm {and} \
\mathbf{M} =
\left( \begin{array}{cccc}
-\Gamma_1 & \Gamma_{21} & \Gamma_{31} & \cdots\\
\Gamma_{12} & -\Gamma_2 & \Gamma_{32} & \cdots\\
\Gamma_{13} & \Gamma_{23} & -\Gamma_{3} &\cdots\\
\vdots & \vdots & \vdots & \ddots \\
\end{array} \right),
\end {equation}
where $\Gamma_i = \sum_{j \ne i} \Gamma_{ij}$ is the rate of
jumping out of site $i$. The equation is then $\mathbf{M}
\mathbf{p} = 0$, which we solve numerically. The matrix
$\mathbf{M}$ defined in this way is singular, which makes a direct
solution rather difficult. By replacing one of the balance
equations with the normalization
\begin {equation}
\sum_i p_i = 1,
\end {equation}
the matrix becomes nonsingular, and the solution can be obtained
more efficiently. Additionally, the solution obtained in this way
is correctly normalized. After this replacement, the equation has
the form:
\begin {equation}
\left( \begin{array}{cccc}
1 & 1 & 1 & \cdots \\
\Gamma_{12} & -\Gamma_2 & \Gamma_{32} & \cdots\\
\Gamma_{13} & \Gamma_{23} & -\Gamma_{3} &\cdots\\
\vdots & \vdots & \vdots & \ddots \\
\end{array} \right)
\left( \begin{array}{c}
p_1 \\
p_2 \\
p_3 \\
\vdots \\
\end{array} \right)
=
\left( \begin{array}{c}
1\\
0\\
0\\
\vdots \\
\end{array} \right)
\end {equation}

As in Section \ref{sec-mc-inf}, the rates for jumps longer than
$d_\text{max}$ are assumed to be zero. Hence, it is efficient to
use a sparse storage scheme for the matrix, where only the
non-zero elements are stored. We obtained the most accurate
results in the shortest time by solving the equation by LU
factorization (in Matlab or Octave with the $\backslash$
operator). This method demands much memory, and did not work for
$L$ above about $22\,R$ on a 32-bit computer.

When the steady-state occupation probabilities are known, the average 
velocity of a charge carrier along the field direction is given by
\begin {equation}
\langle v_x \rangle = \sum_{i, j \ne i} p_i \Gamma_{ij} (x_j -
x_i),
\end {equation}
and the mobility is then $\mu = \frac {\langle v_x \rangle} {F}$.

\subsection {Field dependence of the current density}

\begin{figure}
\includegraphics{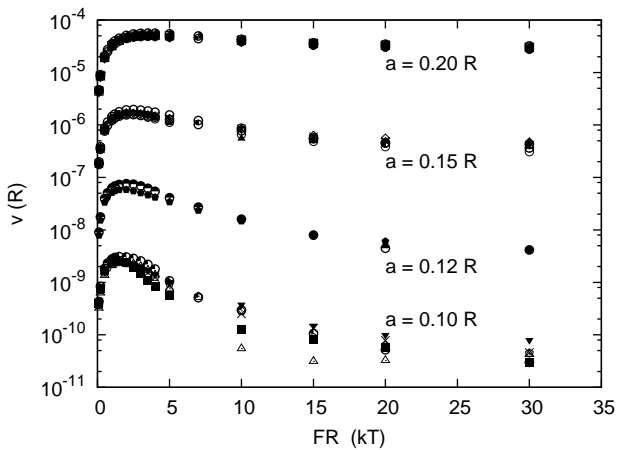}
\caption{Drift velocity as a function of the electric field, for different
localization lengths. The system size $L$ is $20\,R$.}
\label {fig-loclen}
\end{figure}

The simulated dependence of the drift velocity $v$ on the electric
field $F$ is presented in Fig.~\ref{fig-loclen}. Simulations are
performed for $20^3$ sites in a cubic domain with the size 
$L=20\,R$, in the limit
of infinitely small electron concentration. Different symbols
refer to different localization lengths and/or different
realizations of the distribution of sites in the domain.
The size of the simulated system was 10 times larger than 
that in the simulations of Levin et al.,\cite {Shklovskii76r} whose computer 
simulations for the first time confirmed the existence of the NDC effect for hopping transport.

\begin{figure}
\includegraphics{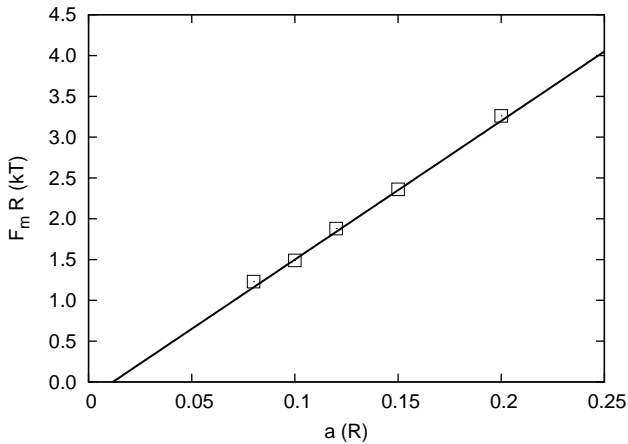}
\caption{The field $F_m$ corresponding to the maximum of the drift
velocity as a function of the localization length $a$. The linear
fit is given by $F_mR/kT = 17 a/R - 0.2$.} \label {fig-peakpos}
\end{figure}

In the limit of small electric field, $FR/kT\ll1$, simulations
show an Ohmic conductivity, i.e., $v$ is  proportional to $F$, in
accordance to the Miller--Abrahams concept of the resistance
network.\cite{Miller1960,bible} With increasing field, the drift
velocity reaches a maximum value. The field strength $F_m$
corresponding to the maximum of the velocity appears to be nearly
proportional to the localization length $a$ within the range
$0.08\,R<a<0.2\,R$ (see Fig.~\ref{fig-peakpos}). At field
strengths $F>F_m$ the NDC appears, i.~e. the drift velocity drops
with increasing field. Simulations show the presence of the NDC
for localization lengths up to $0.3\,R$; when the localization
length is decreased, the NDC effect becomes more pronounced.

\begin{figure}
\includegraphics{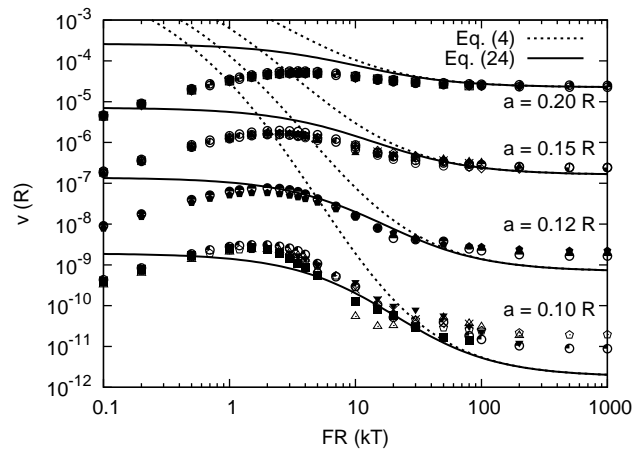}
\caption{Drift velocity as a function of the electric field for
different localization lengths. The curves show Eq.~(\ref
{eq-shklovskii10}) and Eq.~(\ref {eq-shklovskii10-mod}), scaled to
approach Eq.~(\ref{eq-v-improved}) in the limit of large fields.
The system size $L$ is $20\,R$.} \label {fig-loclen-log}
\end{figure}

Fig.~\ref{fig-loclen-log} shows the comparison between the
simulation results (symbols) and the predictions of Nguyen and
Shklovskii\cite{NS} (Eq.~(\ref{eq-shklovskii10}), dashed lines).
For better agreement between the theory and the simulation at
$F\rightarrow\infty$, we take into account the $F$-independent
correction~(\ref{eq-v-improved}) to Eq.~(\ref{eq-shklovskii10}).
Some discrepancies between the simulated and predicted drift
velocities remain at large fields for $a=0.10\,R$ and $a=0.12\,R$.
We believe that these discrepancies are due to the small size of the
simulated system. In fact, the simulated system must be large
enough to contain a reasonable number of optimal traps. The
concentration of optimal traps decreases sharply with decreasing
localization length, so that at smaller localization lengths
larger systems are needed. Thus, for small localization length
($0.10\,R$ and $0.15\,R$), only the shape of the simulated filed
dependence should be taken as representative, but not the values of the
calculated velocities themselves.

The range of applicability of Eq.~(\ref{eq-shklovskii10}) is
determined by the condition $FR/kT\gg1$. One can see nevertheless
that even within this range the field dependence of the velocity
is much weaker than the one predicted by
Eq.~(\ref{eq-shklovskii10}). This result forced us to consider
another possible optimal trap shape for the case of a finite field
as compared to the one considered in Ref. \onlinecite{NS}.

The essential feature of the optimal trap proposed by Nguyen and
Shklovskii (Fig.~\ref{fig-traps}b) is the chain of sites along the
axis of the cone. This chain was introduced in order to provide an
easy path for an electron into a trap. The chain affects the trap
shape and volume, as it serves also as a channel for escaping of
an electron from the trap. We suggest that at moderate
localization lengths ($a\simeq0.1\,R$) traps without such a chain
can also play a significant role. Our next aim is to consider the
shape of traps without a chain of sites and to estimate their
influence on the electron conduction.

\begin{figure}
\includegraphics[width=3.8 cm]{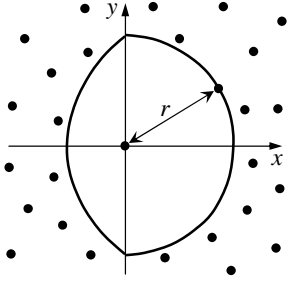}
\caption{Geometry of the optimal trap at finite electric fields
without a chain of sites leading into the trap.} \label
{fig-newtrap}
\end{figure}

A sketch of such a trap is shown in Fig.~\ref{fig-newtrap}. Its
shape is defined by the condition that the rate of jumping from the
central site to any point of the trap's surface is the same. From
Eq.~(\ref{eq-rate}) one can see that in the positive direction
along the axis $X$ the trap is bounded by a hemisphere, and in the
negative direction---by a surface defined by an equation
\begin {equation}
\label{eq-trap-surface}
\frac{F}{kT}\,x-\frac{2}{a}\sqrt{x^2+y^2+z^2} = -\frac{2r}{a},
\end {equation}
where $r$ is the radius of the hemisphere, and the origin is
placed at the central site of the trap. The surface determined by
Eq.~(\ref{eq-trap-surface}) is a quadric surface (an ellipsoid, a
paraboloid, or a hyperboloid, depending on the values of
parameters). The volume of the trap, $V'_\text{trap}$ is
\begin {equation}
\label{eq-trap-volume}
V'_\text{trap} (r) = \frac {2 \pi r^3}{3} 
\left( 1+ \frac{c+2}{2(c+1)^2} \right),
\end {equation}
where $c \equiv Fa/2kT$.

To obtain the drift velocity $v$ (or the current density $j=n_ev$)
in the assumption that the most important traps are those shown in
Fig.~\ref{fig-newtrap}, one can proceed in the same way as the one
applied in Section~\ref{sec-model} to get
Eq.~(\ref{eq-shklovskii10}); the only difference is using
$V'_\text{trap}(r)$ instead of Nguyen and Shklovskii's trap volume
$V_\text{trap} (r) = \frac {2 \pi r^3}{3} \left( 1+ \frac{kT}{Fa}
\right)$. The result is
\begin{equation}
\label{eq-shklovskii10-mod}
 j_{n_e\rightarrow0} \simeq  n_e (a^3R)^\frac14 \exp \! 
\left[\!-\frac{4}{3\sqrt\pi} \left(\frac Ra\right)^\frac32
\left( 1\!+\! \frac{c+2}{2(c\!+\!1)^2}
\right)^{-\frac12}\right]\!.
\end{equation}
For the optimal trap radius one gets
\[
 r_m = \frac{1}{\sqrt{\pi Na}} \left( 1+\frac{c+2}{2(c+1)^2} \right)^{-1/2} .
\]

Since $V'_\text{trap}(r) < V_\text{trap}(r)$, the new trap shape
gives a weaker field dependence of the drift velocity, and a
better agreement with the data from simulations, as one can see in
Fig.~\ref{fig-loclen-log}. However, the simulated field dependence
appears even weaker than the one expressed by
Eq.~(\ref{eq-shklovskii10-mod}). It leads to the assumption that an
actual optimal trap has a shape different from both
Fig.~\ref{fig-traps}b and Fig.~\ref{fig-newtrap}, and hence has a
different volume.

\begin{figure}
\includegraphics{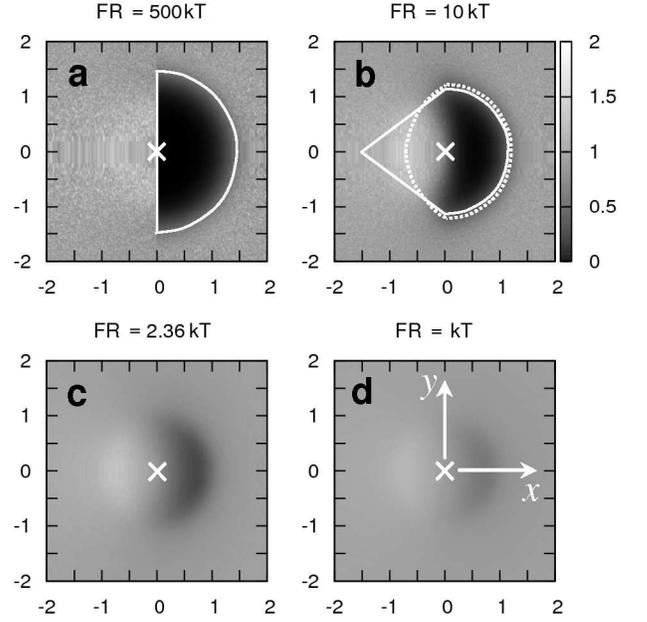}
\caption{Time-averaged density of sites around the charge carrier
for different fields at the localization length $a=0.15\,R$. The
position of a charge carrier (at the origin) is pointed out by a
cross. Boundaries of optimal traps predicted in
Ref.~\onlinecite{NS} (see Fig.~\ref{fig-traps}) are depicted by
solid lines, the boundary of the trap sketched in
Fig.~\ref{fig-newtrap} by the dotted line. Spatial coordinates
are in units of $R$. The value $2.36 kT$ for $FR$ corresponds to the
maximum of the drift velocity.} \label {fig-trapdistr}
\end{figure}

To further investigate the shape of the most efficient traps, we
collect information about the trap shape from the simulations.
Fig.~\ref{fig-trapdistr} show the time-average of the density of
sites around the charge carrier. To calculate this density
$\rho(\mathbf r)$, the space was divided into small elements of
equal volume $\Delta V$; then $\rho(\mathbf r)$ was evaluated as a
sum over pairs of sites:
\[
 \rho(\mathbf r) = \frac{1}{\Delta V} \sum_{i\neq j} p_i\, \chi_{ij}(\mathbf r),
\]
where $p_i$ is an occupation probability of the $i$-th site;
$\chi_{ij}(\mathbf r)=1$ if the vector $(\mathbf r_j - \mathbf
r_i)$ belongs to the same spatial element as the vector $\mathbf
r$; otherwise $\chi_{ij}(\mathbf r)=0$. Finally, values of
$\rho(\mathbf r)$ were averaged over several realizations of site
distributions.

Since the carrier spends the most time in the efficient traps, the
density distribution $\rho(\mathbf r)$ directly reflects the shape
of these traps. At high fields (Fig.~\ref{fig-trapdistr}a) the
hemispherical shape and the size of the trap are in an excellent
agreement with the Nguyen and Shklovskii's theory. However, at
moderate fields, in the region of the NDC
(Fig.~\ref{fig-trapdistr}(b)), neither Fig.~\ref{fig-traps}(b) nor
Fig.~\ref{fig-newtrap} describe the simulated optimal trap. The
optimal trap consists in such a case of a hemisphere in the
spatial region $x>0$, and of a toroidal ``barrier'' in the region
$x<0$, adjoining to a periphery of the hemisphere. The volume of
the optimal trap turns out to be smaller than the one predicted by
both Eq.~(\ref {eq-shklovskii10}) and Eq.~(\ref
{eq-shklovskii10-mod}), in accordance with the result that the
simulated NDC effect is weaker than the predicted one. We would
like to emphasize that the numerically obtained NDC has exactly
the origin predicted by Nguyen and Shklovskii,\cite{NS} consisting
in spreading of the optimal trap into the region $x<0$ at finite
fields and consequently in the increase of the trap volume with
decreasing the field strength.
The possibility of traps in the form of clusters instead of single 
chains of sites has been considered in Ref.~\onlinecite{Shklovskii114r}. 
We interpret our numerical result as a confirmation of that idea. 

A further decrease of the electric field $F$ results in the washing
out the empty region in the density of sites $\rho(\mathbf r)$, as
shown in Fig.~\ref{fig-trapdistr}(c) for $F=F_m$. Finally, at small
$F$ the trap almost disappeared (Fig.~\ref{fig-trapdistr}(d)), which
points to a negligible role of the trapping effect in the regime
of Ohmic conduction.

\begin{figure}
\includegraphics{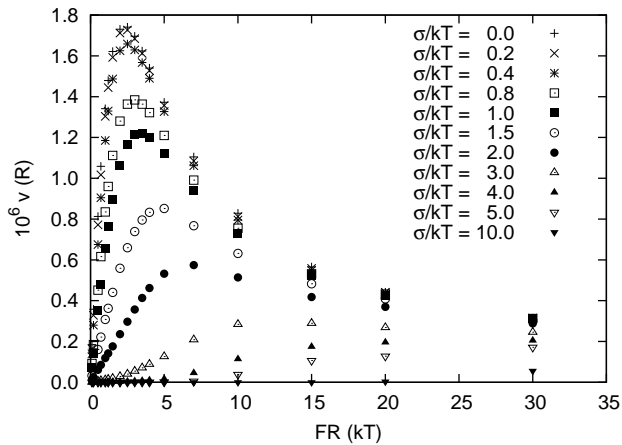}
\caption{Drift velocity $v$ versus electric field $F$ for Gaussian
disorder in site energies. Disorder is characterized by a standard
deviation $\sigma$ of site energies from the reference energy. The
localization length $a = 0.15\,R$.} \label {fig-disorder}
\end{figure}

Materials studied experimentally usually have disorder not only in 
the spatial distribution of localized states, 
but also in the site energies.\cite{Bassler1993,Bassler2000,Rubel04} 
It is therefore necessary to
check how stable the NDC effect is with respect to energetic
disorder. In order to study the role of the energetic disorder for
the NDC effect, we repeated the simulation in a system with a Gaussian
distribution of site energies,  characterized by the standard
deviation $\sigma$.
Fig.~\ref{fig-disorder} shows that introducing a random energy for
each site (with a Gaussian distribution) decreases the drift
velocity and it also decreases the height of the peak of the
velocity as a function of the electric field as compared to
systems with only spatial disorder. 
This weakening of the NDC effect with the increase of the 
energetic disorder (or with the decrease of temperature) is in 
agreement with experimental observations.\cite{Shklovskii114r,Shklovskii116}
Generally, the NDC effect is
confirmed herewith also for systems with the energetic disorder.
However, at extremely large energetic disorder (parameter
$\sigma$), the peak in the field dependence of the drift velocity
disappears completely. The effect of energetic disorder becomes
smaller at larger fields, as expected from the fact that in the
limit $F\rightarrow\infty$ the hopping rates do not depend on site
energies.\\

\section {Conclusions}
\label {sec-conclusions} Numerical studies of the field-dependent
drift velocity of charge carriers in the hopping regime at high
electric fields confirm the prediction of the existing analytical
theories\cite{Bottger1979, NS} that the negative differential
conductivity is inherent for this transport mode. However, the
shape of the field dependence on the current density obtained
numerically differs essentially from the one predicted so
far.\cite{NS} The analytical theory has been improved to give a
much better agreement with the numerical results. In the limit of
the infinitely high electric fields, the predictions of the
analytical theory of Nguyen and Shklovskii\cite{NS} are to much
extent confirmed by our straightforward Monte Carlo simulations.

\begin{acknowledgments}
The authors are indebted to Prof. Boris Shklovskii for numerous valuable comments. 
Financial support from the Academy of Finland project 116995 and
the TEKES NAMU project, from the Deutsche Forschungsgemeinschaft and
that of the Fonds der Che\-mischen Industrie is gratefully
acknowledged.
A.~V.~N. thanks the Russian Foundation for Basic Research (project 06-02-16988)
and the Dynasty Foundation for financial support.
The authors thank Oleg Rubel and Kakhaber Jandieri for stimulating discussions.

\end{acknowledgments}


\begin{thebibliography}{39}
\expandafter\ifx\csname natexlab\endcsname\relax\def\natexlab#1{#1}\fi
\expandafter\ifx\csname bibnamefont\endcsname\relax
  \def\bibnamefont#1{#1}\fi
\expandafter\ifx\csname bibfnamefont\endcsname\relax
  \def\bibfnamefont#1{#1}\fi
\expandafter\ifx\csname citenamefont\endcsname\relax
  \def\citenamefont#1{#1}\fi
\expandafter\ifx\csname url\endcsname\relax
  \def\url#1{\texttt{#1}}\fi
\expandafter\ifx\csname urlprefix\endcsname\relax\def\urlprefix{URL }\fi
\providecommand{\bibinfo}[2]{#2}
\providecommand{\eprint}[2][]{\url{#2}}

\bibitem[{\citenamefont{B\"{o}ttger and Bryksin}(1985)}]{Bottger1985}
\bibinfo{author}{\bibfnamefont{H.}~\bibnamefont{B\"{o}ttger}} \bibnamefont{and}
  \bibinfo{author}{\bibfnamefont{V.~V.} \bibnamefont{Bryksin}},
  \emph{\bibinfo{title}{Hopping conduction in solids}} (\bibinfo{publisher}{VCH
  Akademie-Verlag Berlin}, \bibinfo{year}{1985}).

\bibitem[{\citenamefont{Baranovski}(2006)}]{Baranovski2006}
\bibinfo{editor}{\bibfnamefont{S.}~\bibnamefont{Baranovski}}, ed.,
  \emph{\bibinfo{title}{Charge Transport in Disordered Solids with Applications
  in Electronics}} (\bibinfo{publisher}{John Wiley \& Sons, Ltd, Chichester},
  \bibinfo{year}{2006}).

\bibitem[{\citenamefont{B\"{a}ssler}(2000)}]{Bassler2000}
\bibinfo{author}{\bibfnamefont{H.}~\bibnamefont{B\"{a}ssler}},
  \emph{\bibinfo{title}{Semiconducting Polymers, G. Hadziioannou and P. F. van
  Hutten (eds.)}} (\bibinfo{publisher}{John Wiley \& Sons, Inc., New York},
  \bibinfo{year}{2000}), p. \bibinfo{pages}{365}.

\bibitem[{\citenamefont{Pope and Swenberg}(1999)}]{Pope1999}
\bibinfo{author}{\bibfnamefont{M.}~\bibnamefont{Pope}} \bibnamefont{and}
  \bibinfo{author}{\bibfnamefont{C.~E.} \bibnamefont{Swenberg}},
  \emph{\bibinfo{title}{Electronic Processes in Organic Crystals and Polymers}}
  (\bibinfo{publisher}{Oxford University Press, Oxford}, \bibinfo{year}{1999}).

\bibitem[{\citenamefont{B\"assler}(1993)}]{Bassler1993}
\bibinfo{author}{\bibfnamefont{H.}~\bibnamefont{B\"assler}},
  \bibinfo{journal}{Phys. Status Solidi B} \textbf{\bibinfo{volume}{175}},
  \bibinfo{pages}{15} (\bibinfo{year}{1993}).

\bibitem[{\citenamefont{Borsenberger et~al.}(1993)\citenamefont{Borsenberger,
  Magin, van~der Auweraer, and de~Schryver}}]{Borsenberger1993}
\bibinfo{author}{\bibfnamefont{P.~M.} \bibnamefont{Borsenberger}},
  \bibinfo{author}{\bibfnamefont{E.~H.} \bibnamefont{Magin}},
  \bibinfo{author}{\bibfnamefont{M.}~\bibnamefont{van~der Auweraer}},
  \bibnamefont{and} \bibinfo{author}{\bibfnamefont{F.~C.}
  \bibnamefont{de~Schryver}}, \bibinfo{journal}{Phys. Status Solidi (a)}
  \textbf{\bibinfo{volume}{140}}, \bibinfo{pages}{9} (\bibinfo{year}{1993}).

\bibitem[{\citenamefont{van~der Auweraer et~al.}(1994)\citenamefont{van~der
  Auweraer, de~Schryver, Borsenberger, and B\"assler}}]{Auweraer1994}
\bibinfo{author}{\bibfnamefont{M.}~\bibnamefont{van~der Auweraer}},
  \bibinfo{author}{\bibfnamefont{F.~C.} \bibnamefont{de~Schryver}},
  \bibinfo{author}{\bibfnamefont{P.~M.} \bibnamefont{Borsenberger}},
  \bibnamefont{and}
  \bibinfo{author}{\bibfnamefont{H.}~\bibnamefont{B\"assler}},
  \bibinfo{journal}{Advanced Materials} \textbf{\bibinfo{volume}{6}},
  \bibinfo{pages}{199} (\bibinfo{year}{1994}).

\bibitem[{\citenamefont{Abkowitz}(1992)}]{Abkowitz1992}
\bibinfo{author}{\bibfnamefont{M.}~\bibnamefont{Abkowitz}},
  \bibinfo{journal}{Phil. Mag. B} \textbf{\bibinfo{volume}{65}},
  \bibinfo{pages}{817} (\bibinfo{year}{1992}).

\bibitem[{\citenamefont{Peled and Schein}(1988)}]{Peled1988}
\bibinfo{author}{\bibfnamefont{A.}~\bibnamefont{Peled}} \bibnamefont{and}
  \bibinfo{author}{\bibfnamefont{L.~B.} \bibnamefont{Schein}},
  \bibinfo{journal}{Chem. Phys. Lett.} \textbf{\bibinfo{volume}{153}},
  \bibinfo{pages}{422} (\bibinfo{year}{1988}).

\bibitem[{\citenamefont{Schein}(1992)}]{Schein1992}
\bibinfo{author}{\bibfnamefont{L.~B.} \bibnamefont{Schein}},
  \bibinfo{journal}{Phil. Mag. B} \textbf{\bibinfo{volume}{65}},
  \bibinfo{pages}{795} (\bibinfo{year}{1992}).

\bibitem[{\citenamefont{Parris and Bookout}(1996)}]{Parris96}
\bibinfo{author}{\bibfnamefont{P.~E.} \bibnamefont{Parris}} \bibnamefont{and}
  \bibinfo{author}{\bibfnamefont{B.~D.} \bibnamefont{Bookout}},
  \bibinfo{journal}{Phys. Rev. B} \textbf{\bibinfo{volume}{53}},
  \bibinfo{pages}{629} (\bibinfo{year}{1996}).

\bibitem[{\citenamefont{Hirao et~al.}(1995)\citenamefont{Hirao, Nishizawa, and
  Sugiuchi}}]{Hirao1995}
\bibinfo{author}{\bibfnamefont{A.}~\bibnamefont{Hirao}},
  \bibinfo{author}{\bibfnamefont{H.}~\bibnamefont{Nishizawa}},
  \bibnamefont{and} \bibinfo{author}{\bibfnamefont{M.}~\bibnamefont{Sugiuchi}},
  \bibinfo{journal}{Phys. Rev. Lett.} \textbf{\bibinfo{volume}{75}},
  \bibinfo{pages}{1787} (\bibinfo{year}{1995}).

\bibitem[{\citenamefont{Cordes et~al.}(2001)\citenamefont{Cordes, Baranovskii,
  Kohary, Thomas, Yamasaki, Hensel, and Wendorff}}]{Cordes2001}
\bibinfo{author}{\bibfnamefont{H.}~\bibnamefont{Cordes}},
  \bibinfo{author}{\bibfnamefont{S.~D.} \bibnamefont{Baranovskii}},
  \bibinfo{author}{\bibfnamefont{K.}~\bibnamefont{Kohary}},
  \bibinfo{author}{\bibfnamefont{P.}~\bibnamefont{Thomas}},
  \bibinfo{author}{\bibfnamefont{S.}~\bibnamefont{Yamasaki}},
  \bibinfo{author}{\bibfnamefont{F.}~\bibnamefont{Hensel}}, \bibnamefont{and}
  \bibinfo{author}{\bibfnamefont{J.-H.} \bibnamefont{Wendorff}},
  \bibinfo{journal}{Phys. Rev. B} \textbf{\bibinfo{volume}{63}},
  \bibinfo{pages}{094201} (\bibinfo{year}{2001}).

\bibitem[{\citenamefont{B\"ottger and Bryksin}(1979)}]{Bottger1979}
\bibinfo{author}{\bibfnamefont{H.}~\bibnamefont{B\"ottger}} \bibnamefont{and}
  \bibinfo{author}{\bibfnamefont{V.~V.} \bibnamefont{Bryksin}},
  \bibinfo{journal}{Phys. Status Solidi B} \textbf{\bibinfo{volume}{96}},
  \bibinfo{pages}{219} (\bibinfo{year}{1979}).

\bibitem[{\citenamefont{{Nguyen Van Lien} and Shklovskii}(1981)}]{NS}
\bibinfo{author}{\bibnamefont{{Nguyen Van Lien}}} \bibnamefont{and}
  \bibinfo{author}{\bibfnamefont{B.~I.} \bibnamefont{Shklovskii}},
  \bibinfo{journal}{Solid State Commun.} \textbf{\bibinfo{volume}{38}},
  \bibinfo{pages}{99} (\bibinfo{year}{1981}).

\bibitem[{\citenamefont{Levin and Shklovskii}(1988)}]{Levin1988}
\bibinfo{author}{\bibfnamefont{E.~I.} \bibnamefont{Levin}} \bibnamefont{and}
  \bibinfo{author}{\bibfnamefont{B.~I.} \bibnamefont{Shklovskii}},
  \bibinfo{journal}{Solid State Commun.} \textbf{\bibinfo{volume}{67}},
  \bibinfo{pages}{233} (\bibinfo{year}{1988}).

\bibitem[{\citenamefont{Aladashvili
  et~al.}(1988{\natexlab{a}})\citenamefont{Aladashvili, Adamia, Lavdovskii,
  Levin, and Shklovskii}}]{Shklovskii99r}
\bibinfo{author}{\bibfnamefont{D.~I.} \bibnamefont{Aladashvili}},
  \bibinfo{author}{\bibfnamefont{Z.~A.}~\bibnamefont{Adamia}},
  \bibinfo{author}{\bibfnamefont{K.~G.}~\bibnamefont{Lavdovskii}},
  \bibinfo{author}{\bibfnamefont{E.~I.} \bibnamefont{Levin}}, \bibnamefont{and}
  \bibinfo{author}{\bibfnamefont{B.~I.} \bibnamefont{Shklovskii}},
  \bibinfo{journal}{Pis'ma v Zh. Eksp. Teor. Fiz.}
  \textbf{\bibinfo{volume}{47}}, \bibinfo{pages}{390} (\bibinfo{year}{1988}{\natexlab{a}}),
  \bibinfo{journal}{Sov. Phys. JETP Lett.} \textbf{\bibinfo{volume}{47}}, \bibinfo{pages}{466}
  (\bibinfo{year}{1988}{\natexlab{b}}).

%

\bibitem[{\citenamefont{Aladashvili
  et~al.}(1990{\natexlab{a}})\citenamefont{Aladashvili, Adamia, Lavdovskii,
  Levin, and Shklovskii}}]{Shklovskii114r}
\bibinfo{author}{\bibfnamefont{D.~I.} \bibnamefont{Aladashvili}},
  \bibinfo{author}{\bibfnamefont{Z.~A.}~\bibnamefont{Adamia}},
  \bibinfo{author}{\bibfnamefont{K.~G.}~\bibnamefont{Lavdovskii}},
  \bibinfo{author}{\bibfnamefont{E.~I.} \bibnamefont{Levin}}, \bibnamefont{and}
  \bibinfo{author}{\bibfnamefont{B.~I.} \bibnamefont{Shklovskii}},
  \bibinfo{journal}{Fiz. Tekhn. Poluprov.} \textbf{\bibinfo{volume}{24}}, \bibinfo{pages}{234}
  (\bibinfo{year}{1990}{\natexlab{a}}),  \bibinfo{journal}{Sov. Phys. Semicond.} \textbf{\bibinfo{volume}{24}}, \bibinfo{pages}{143} (\bibinfo{year}{1990}{\natexlab{b}}).


%

\bibitem[{\citenamefont{Aladashvili
  et~al.}(1990{\natexlab{c}})\citenamefont{Aladashvili, Adamiya, Lavdovskii,
  Levin, and Shklovskii}}]{Shklovskii116}
\bibinfo{author}{\bibfnamefont{D.~I.} \bibnamefont{Aladashvili}},
  \bibinfo{author}{\bibfnamefont{Z.~A.} \bibnamefont{Adamiya}},
  \bibinfo{author}{\bibfnamefont{K.~G.} \bibnamefont{Lavdovskii}},
  \bibinfo{author}{\bibfnamefont{E.~I.} \bibnamefont{Levin}}, \bibnamefont{and}
  \bibinfo{author}{\bibfnamefont{B.~I.} \bibnamefont{Shklovskii}}, in
  \emph{\bibinfo{booktitle}{Hopping and related phenomena}}, edited by
  \bibinfo{editor}{\bibfnamefont{H.}~\bibnamefont{Fritzsche}} \bibnamefont{and}
  \bibinfo{editor}{\bibfnamefont{M.}~\bibnamefont{Pollak}}
  (\bibinfo{publisher}{World Scientific}, \bibinfo{year}{1990}{\natexlab{c}}).

\bibitem[{\citenamefont{Shklovskii and Efros}(1984)}]{bible}
\bibinfo{author}{\bibfnamefont{B.~I.} \bibnamefont{Shklovskii}}
  \bibnamefont{and} \bibinfo{author}{\bibfnamefont{A.~L.} \bibnamefont{Efros}},
  \emph{\bibinfo{title}{Electronic Properties of Doped Semiconductors}}
  (\bibinfo{publisher}{Springer-Verlag}, \bibinfo{year}{1984}).

\bibitem[{\citenamefont{Levin et~al.}(1982{\natexlab{a}})\citenamefont{Levin,
  {Nguyen Van Lien}, and Shklovskii}}]{Shklovskii76r}
\bibinfo{author}{\bibfnamefont{E.~I.} \bibnamefont{Levin}},
  \bibinfo{author}{\bibnamefont{{Nguyen Van Lien}}}, \bibnamefont{and}
  \bibinfo{author}{\bibfnamefont{B.~I.} \bibnamefont{Shklovskii}},
  \bibinfo{journal}{Fiz. Tekh. Poluprov} \textbf{\bibinfo{volume}{16}}, \bibinfo{pages}{815}
  (\bibinfo{year}{1982}{\natexlab{a}}),
  \bibinfo{journal}{Sov. Phys. Semicond.} \textbf{\bibinfo{volume}{16}}, \bibinfo{pages}{523}
  (\bibinfo{year}{1982}{\natexlab{b}}).

%

\bibitem[{\citenamefont{Simmons and Verderber}(1967)}]{Simmons1967}
\bibinfo{author}{\bibfnamefont{J.~G.} \bibnamefont{Simmons}} \bibnamefont{and}
  \bibinfo{author}{\bibfnamefont{R.~R.} \bibnamefont{Verderber}},
  \bibinfo{journal}{Proc. R. Soc. London A} \textbf{\bibinfo{volume}{391}},
  \bibinfo{pages}{77} (\bibinfo{year}{1967}).

\bibitem[{\citenamefont{Thurstans and Oxley}(2002)}]{Thurstans2002}
\bibinfo{author}{\bibfnamefont{R.~E.} \bibnamefont{Thurstans}}
  \bibnamefont{and} \bibinfo{author}{\bibfnamefont{D.~P.} \bibnamefont{Oxley}},
  \bibinfo{journal}{J. Phys. D} \textbf{\bibinfo{volume}{35}},
  \bibinfo{pages}{802} (\bibinfo{year}{2002}).

\bibitem[{\citenamefont{Bozano et~al.}(2004)\citenamefont{Bozano, Kean, Deline,
  Salem, and Scott}}]{Bozano2004}
\bibinfo{author}{\bibfnamefont{L.~D.} \bibnamefont{Bozano}},
  \bibinfo{author}{\bibfnamefont{B.~W.} \bibnamefont{Kean}},
  \bibinfo{author}{\bibfnamefont{V.~R.} \bibnamefont{Deline}},
  \bibinfo{author}{\bibfnamefont{J.~R.} \bibnamefont{Salem}}, \bibnamefont{and}
  \bibinfo{author}{\bibfnamefont{J.~C.} \bibnamefont{Scott}},
  \bibinfo{journal}{Appl. Phys. Lett.} \textbf{\bibinfo{volume}{84}},
  \bibinfo{pages}{607} (\bibinfo{year}{2004}).

\bibitem[{\citenamefont{Bozano et~al.}(2005)\citenamefont{Bozano, Kean,
  Beinhoff, Carter, Rice, and Scott}}]{Bozano2005}
\bibinfo{author}{\bibfnamefont{L.~D.} \bibnamefont{Bozano}},
  \bibinfo{author}{\bibfnamefont{B.~W.} \bibnamefont{Kean}},
  \bibinfo{author}{\bibfnamefont{M.}~\bibnamefont{Beinhoff}},
  \bibinfo{author}{\bibfnamefont{K.~R.} \bibnamefont{Carter}},
  \bibinfo{author}{\bibfnamefont{P.~M.} \bibnamefont{Rice}}, \bibnamefont{and}
  \bibinfo{author}{\bibfnamefont{J.~C.} \bibnamefont{Scott}},
  \bibinfo{journal}{Advanced Functional Materials}
  \textbf{\bibinfo{volume}{15}}, \bibinfo{pages}{1933} (\bibinfo{year}{2005}).

\bibitem[{\citenamefont{Majumdar et~al.}(2005)\citenamefont{Majumdar, Baral,
  \"Osterbacka, Ikkala, and Stubb}}]{Majumdar2005}
\bibinfo{author}{\bibfnamefont{H.~S.} \bibnamefont{Majumdar}},
  \bibinfo{author}{\bibfnamefont{J.~K.} \bibnamefont{Baral}},
  \bibinfo{author}{\bibfnamefont{R.}~\bibnamefont{\"Osterbacka}},
  \bibinfo{author}{\bibfnamefont{O.}~\bibnamefont{Ikkala}}, \bibnamefont{and}
  \bibinfo{author}{\bibfnamefont{H.}~\bibnamefont{Stubb}},
  \bibinfo{journal}{Organic Electronics} \textbf{\bibinfo{volume}{6}},
  \bibinfo{pages}{188} (\bibinfo{year}{2005}).

\bibitem[{\citenamefont{Verbakel et~al.}(2007)\citenamefont{Verbakel, Meskers,
  Janssen, Gomes, Colle, Buchel, and de~Leeuw}}]{Verbakel2007}
\bibinfo{author}{\bibfnamefont{F.}~\bibnamefont{Verbakel}},
  \bibinfo{author}{\bibfnamefont{S.~C.~J.} \bibnamefont{Meskers}},
  \bibinfo{author}{\bibfnamefont{R.~A.~J.} \bibnamefont{Janssen}},
  \bibinfo{author}{\bibfnamefont{H.~L.} \bibnamefont{Gomes}},
  \bibinfo{author}{\bibfnamefont{M.}~\bibnamefont{Colle}},
  \bibinfo{author}{\bibfnamefont{M.}~\bibnamefont{Buchel}}, \bibnamefont{and}
  \bibinfo{author}{\bibfnamefont{D.~M.} \bibnamefont{de~Leeuw}},
  \bibinfo{journal}{Appl. Phys. Lett.} \textbf{\bibinfo{volume}{91}},
  \bibinfo{eid}{192103} (\bibinfo{year}{2007}).

\bibitem[{\citenamefont{Baral et~al.}(2008)\citenamefont{Baral, Majumdar,
  Laiho, Jiang, Kauppinen, Ras, Ruokolainen, Ikkala, and
  \"Osterbacka}}]{Baral2008}
\bibinfo{author}{\bibfnamefont{J.~K.} \bibnamefont{Baral}},
  \bibinfo{author}{\bibfnamefont{H.~S.} \bibnamefont{Majumdar}},
  \bibinfo{author}{\bibfnamefont{A.}~\bibnamefont{Laiho}},
  \bibinfo{author}{\bibfnamefont{H.}~\bibnamefont{Jiang}},
  \bibinfo{author}{\bibfnamefont{E.~I.} \bibnamefont{Kauppinen}},
  \bibinfo{author}{\bibfnamefont{R.~H.~A.} \bibnamefont{Ras}},
  \bibinfo{author}{\bibfnamefont{J.}~\bibnamefont{Ruokolainen}},
  \bibinfo{author}{\bibfnamefont{O.}~\bibnamefont{Ikkala}}, \bibnamefont{and}
  \bibinfo{author}{\bibfnamefont{R.}~\bibnamefont{\"Osterbacka}},
  \bibinfo{journal}{Nanotechnology} \textbf{\bibinfo{volume}{19}},
  \bibinfo{pages}{035203} (\bibinfo{year}{2008}).

\bibitem[{\citenamefont{Yu et~al.}(2000)\citenamefont{Yu, Smith, Saxena,
  Martin, and Bishop}}]{Yu2000}
\bibinfo{author}{\bibfnamefont{Z.~G.} \bibnamefont{Yu}},
  \bibinfo{author}{\bibfnamefont{D.~L.} \bibnamefont{Smith}},
  \bibinfo{author}{\bibfnamefont{A.}~\bibnamefont{Saxena}},
  \bibinfo{author}{\bibfnamefont{R.~L.} \bibnamefont{Martin}},
  \bibnamefont{and} \bibinfo{author}{\bibfnamefont{A.~R.}
  \bibnamefont{Bishop}}, \bibinfo{journal}{Phys. Rev. Lett.}
  \textbf{\bibinfo{volume}{84}}, \bibinfo{pages}{721} (\bibinfo{year}{2000}).

\bibitem[{\citenamefont{Yu et~al.}(2001)\citenamefont{Yu, Smith, Saxena,
  Martin, and Bishop}}]{Yu2001}
\bibinfo{author}{\bibfnamefont{Z.~G.} \bibnamefont{Yu}},
  \bibinfo{author}{\bibfnamefont{D.~L.} \bibnamefont{Smith}},
  \bibinfo{author}{\bibfnamefont{A.}~\bibnamefont{Saxena}},
  \bibinfo{author}{\bibfnamefont{R.~L.} \bibnamefont{Martin}},
  \bibnamefont{and} \bibinfo{author}{\bibfnamefont{A.~R.}
  \bibnamefont{Bishop}}, \bibinfo{journal}{Phys. Rev. B}
  \textbf{\bibinfo{volume}{63}}, \bibinfo{pages}{085202}
  (\bibinfo{year}{2001}).

\bibitem[{\citenamefont{Pasveer et~al.}(2005)\citenamefont{Pasveer, Cottaar,
  Bobbert, and Michels}}]{Pasveer2005}
\bibinfo{author}{\bibfnamefont{W.~F.} \bibnamefont{Pasveer}},
  \bibinfo{author}{\bibfnamefont{J.}~\bibnamefont{Cottaar}},
  \bibinfo{author}{\bibfnamefont{P.~A.} \bibnamefont{Bobbert}},
  \bibnamefont{and} \bibinfo{author}{\bibfnamefont{M.~A.~J.}
  \bibnamefont{Michels}}, \bibinfo{journal}{Synth. Met.}
  \textbf{\bibinfo{volume}{152}}, \bibinfo{pages}{157} (\bibinfo{year}{2005}).

\bibitem[{\citenamefont{Cottaar and Bobbert}(2006)}]{Cottaar2006}
\bibinfo{author}{\bibfnamefont{J.}~\bibnamefont{Cottaar}} \bibnamefont{and}
  \bibinfo{author}{\bibfnamefont{P.~A.} \bibnamefont{Bobbert}},
  \bibinfo{journal}{Phys. Rev. B} \textbf{\bibinfo{volume}{74}},
  \bibinfo{eid}{115204} (\bibinfo{year}{2006}).

\bibitem[{\citenamefont{Jansson
  et~al.}(2008{\natexlab{a}})\citenamefont{Jansson, Baranovskii, Sliau\v{z}ys,
  \"Osterbacka, and Thomas}}]{Jansson2008}
\bibinfo{author}{\bibfnamefont{F.}~\bibnamefont{Jansson}},
  \bibinfo{author}{\bibfnamefont{S.~D.} \bibnamefont{Baranovskii}},
  \bibinfo{author}{\bibfnamefont{G.}~\bibnamefont{Sliau\v{z}ys}},
  \bibinfo{author}{\bibfnamefont{R.}~\bibnamefont{\"Osterbacka}},
  \bibnamefont{and} \bibinfo{author}{\bibfnamefont{P.}~\bibnamefont{Thomas}},
  \bibinfo{journal}{Phys. Status Solidi C} \textbf{\bibinfo{volume}{5}},
  \bibinfo{pages}{722} (\bibinfo{year}{2008}{\natexlab{a}}).

\bibitem[{\citenamefont{Jansson
  et~al.}(2008{\natexlab{b}})\citenamefont{Jansson, Baranovskii, Gebhard, and
  \"{O}sterbacka}}]{Jansson2008b}
\bibinfo{author}{\bibfnamefont{F.}~\bibnamefont{Jansson}},
  \bibinfo{author}{\bibfnamefont{S.~D.} \bibnamefont{Baranovskii}},
  \bibinfo{author}{\bibfnamefont{F.}~\bibnamefont{Gebhard}}, \bibnamefont{and}
  \bibinfo{author}{\bibfnamefont{R.}~\bibnamefont{\"{O}sterbacka}},
  \bibinfo{journal}{Phys. Rev. B} \textbf{\bibinfo{volume}{77}},
  \bibinfo{eid}{195211} (\bibinfo{year}{2008}{\natexlab{b}}).

\bibitem[{\citenamefont{Miller and Abrahams}(1960)}]{Miller1960}
\bibinfo{author}{\bibfnamefont{A.}~\bibnamefont{Miller}} \bibnamefont{and}
  \bibinfo{author}{\bibfnamefont{E.}~\bibnamefont{Abrahams}},
  \bibinfo{journal}{Phys. Rev.} \textbf{\bibinfo{volume}{120}},
  \bibinfo{pages}{745} (\bibinfo{year}{1960}).

\bibitem[{\citenamefont{Rubel et~al.}(2004)\citenamefont{Rubel, Baranovskii,
  Thomas, and Yamasaki}}]{Rubel04}
\bibinfo{author}{\bibfnamefont{O.}~\bibnamefont{Rubel}},
  \bibinfo{author}{\bibfnamefont{S.~D.} \bibnamefont{Baranovskii}},
  \bibinfo{author}{\bibfnamefont{P.}~\bibnamefont{Thomas}}, \bibnamefont{and}
  \bibinfo{author}{\bibfnamefont{S.}~\bibnamefont{Yamasaki}},
  \bibinfo{journal}{Phys. Rev. B} \textbf{\bibinfo{volume}{69}},
  \bibinfo{pages}{014206} (\bibinfo{year}{2004}).

\end{thebibliography}

\end{document}